\documentclass[11pt,twoside]{article}

\def\mearth{{\rm\,M_\oplus}}

\def\msun{{\rm\,M_\odot}}

\def\gsim{\rlap{$>$}{\lower 1.0ex\hbox{$\sim$}}}
\def\lsim{\rlap{$<$}{\lower 1.0ex\hbox{$\sim$}}}
\def\circdot{\rlap{$^\circ$}{\hbox{.}}}
\def\etal{{\it et al.\thinspace}}
\def\wpm2{W m$^{-2}$}
\def\etal{{\it et al.\thinspace}}

\def\ie{{\it i.e.\ }}

\usepackage{asp2006}
\usepackage{epsf}
\usepackage{psfig}
\usepackage{lscape}

\begin{document}
\title{Tidal Constraints on Planetary Habitability}   %%% Fill in title
\author{Rory Barnes\altaffilmark{1,2}, Brian
  Jackson\altaffilmark{3,4}, Richard Greenberg\altaffilmark{5}, Sean
  N. Raymond\altaffilmark{2,6}, and Ren\'e Heller\altaffilmark{7}}   %%% Fill in author names
\altaffiltext{1}{Department of Astronomy, University of Washington,
  Seattle, WA, 98195-1580}
\altaffiltext{2}{Virtual Planetary Lab}
\altaffiltext{3}{Planetary Systems Laboratory, Goddard Space Flight
Center, Code 693, Greenbelt, MD 20771}
\altaffiltext{4}{NASA Postdoctoral Program Fellow}
\altaffiltext{5}{Lunar and Planetary Laboratory, University of Arizona, Tucson, AZ 85721}
\altaffiltext{6}{Center for Astrophysics and Space Astronomy, University of
 Colorado, UCB 389, Boulder CO 80309-0389}
\altaffiltext{7}{Hamburger Sternwarte, University of Hamburg,
  Gojenbergsweg 112, 21029 Hamburg, Germany}    %%% Fill in author affiliations

\begin{abstract} %%% Abstract to run on from here.
We review how tides may impact the habitability of terrestrial-like
planets. If such planets form around low-mass stars, then planets in
the circumstellar habitable zone will be close enough to their host stars
to experience strong tidal forces. We discuss 1) decay of semi-major axis, 
2) circularization of
eccentric orbits, 3) evolution toward zero obliquity, 4) fixed
rotation rates (not necessarily synchronous), and 5) internal
heating. We briefly describe these effects using the example of a 0.25
$\msun$ star with a 10 $\mearth$ companion. We suggest that the
concept of a habitable zone should be modified to include the effects
of tides.
\end{abstract}

%%% MAIN BODY OF TEXT GOES HERE. CONSULT "INSTRUCTIONS FOR AUTHORS USING
%%% LATEX2E MARKUP", SECTIONS 2.3-2.6 FOR HELP WITH EQUATIONS, FIGURES,
%%% AND TABLES.

\section{Introduction}
Exoplanet surfaces are probably the best places to look for life
beyond the Solar System. Remote sensing of these bodies is still in
its infancy, and the technology does
not yet exist to measure the properties of terrestrial exoplanet
atmospheres directly. Indeed, the scale and precision of the
engineering required to do so is breathtaking. Given these
limitations, a reliable model of habitability is essential in order
to maximize the scientific return of future ground- and space-based
missions with the capability to remotely detect exoplanet atmospheres.

Here we review one often misunderstood issue: The effect of tides. If
the distance between a star and planet is small, $\lsim 0.1$ AU, the
shape of the planet (and star) can become significantly
non-spherical. This asymmetry can change the planet's orbital motion
from that of spherical planets. Simulating the deviations from the
spherical approximation is difficult and uncertain as observations of
the Solar System, binary stars and exoplanets do not yet provide
enough information to distinguish between models. Without firm
constraints, qualitatively different models of the planetary response
to tides exist. The
two most prominent descriptions are the ``constant-phase-lag'' and
``constant-time-lag'' models (Greenberg 2009). In the former, the
tidal bulge is assumed to lag the perturber by a fixed angle, but in
the latter it lags by a fixed time interval. Depending on the
mathematical extension in terms of $e$, the two models may diverge
significantly when $e \gsim 0.3.$ Throughout this review the reader
should remember that the presented magnitudes of tidal effects are
model-dependent. For more on these differences and the details of tidal
models, the reader is referred to recent reviews by Ferraz-Mello \etal
(2008) and Heller \etal (2009).
 
We consider tidal effects in the habitable zone
(HZ) model proposed by Barnes \etal (2008) which utilizes the 50\%
cloud cover HZ of Selsis \etal (2007), but assumes that the orbit
averaged flux determines surface temperature (Williams and Pollard
2002). We use the example of a 10 $\mearth$
planet orbiting a 0.25 $\msun$ star. This choice is arbitrary, but we
note that large terrestrial planets orbiting small stars will be
preferentially discovered by current detection techniques. This
chapter is organized as follows: In $\S$ 2 we discuss orbital
evolution, in $\S$ 3 we describe rotation rates, in $\S$ 4 we consider
the obliquity, and in $\S$ 5 we examine tidal heating.

\section{Orbital Evolution}
Orbital evolution due to tides should be considered
for any potentially habitable world. The asymmetry of the tidal bulge
leads to torques which transfer angular momentum between rotation and
orbits, and the constant flexing of the planet's figure between pericenter 
and
apocenter dissipates energy inside the planet. These two
effects act to circularize and shrink most orbits. In the
constant-phase-lag model, the orbits of close-in exoplanets evolve in
the following way (Goldreich and Soter 1966; see also Jackson \etal 2009):
\begin{equation}
\label{eq:adot}
\frac{da}{dt} = -\Big(\frac{63}{2}\frac{\sqrt{GM_*^3}R_p^5}{m_pQ'_p}e^2 + \frac{9}{2}\frac{\sqrt{G/M_*}R_*^5m_p}{Q'_*}\Big[1+\frac{57}{4}e^2])a^{-11/2}
\end{equation}
\begin{equation}
\label{eq:edot}
\frac{de}{dt} = -\Big(\frac{63}{4}\frac{\sqrt{GM_*^3}R_p^5}{m_pQ'_p} + \frac{225}{16}\frac{\sqrt{G/M_*}R_*^5m_p}{Q'_*}\Big)a^{-13/2}e,
\end{equation}
where $a$ is semi-major axis, $G$ is Newton's gravitational constant,
$m_p$ is the mass of the planet, $Q'_p$ is the planet's tidal
dissipation function divided by two-thirds its Love number, $Q'_*$ is
the star's tidal dissipation function divided by two-thirds its Love
number, $R_p$ is the planet's radius, and $R_*$ is the stellar
radius. The $Q'$ values represent the body's response to tidal
processes and combines a myriad of internal properties, such as
density, equation of state, etc. It is a difficult quantity to
measure, so here we use the standard values of $Q'_* =
10^6$ and $Q'_p = 500$ (Mathieu 1994; Mardling and Lin 2002; Jackson
\etal 2008a). The first terms in Eqs.~(\ref{eq:adot} -- \ref{eq:edot})
represent the effects of the tide raised on the planet, the second the
tide raised on the star.

Eqs.~(\ref{eq:adot} -- \ref{eq:edot}) predict $a$ and $e$ decay with
time. As tides slowly change a planet's orbit, the planet may move out
(through the inner edge) of the habitable zone (HZ). This
possibility was considered in Barnes \etal (2008), who showed for some
example cases the time for a planet to pass through the
inner edge of the HZ. Such sterilizing evolution is most likely to
occur for planets with initially large eccentricity near the inner
edge of the HZ of low mass stars ($\lsim 0.3 \msun$). Even if a planet
does not leave the HZ, the circularization of its orbit can require
billions of years, potentially affecting the climatic evolution of the
planet.

\begin{figure}[t]
\plotone{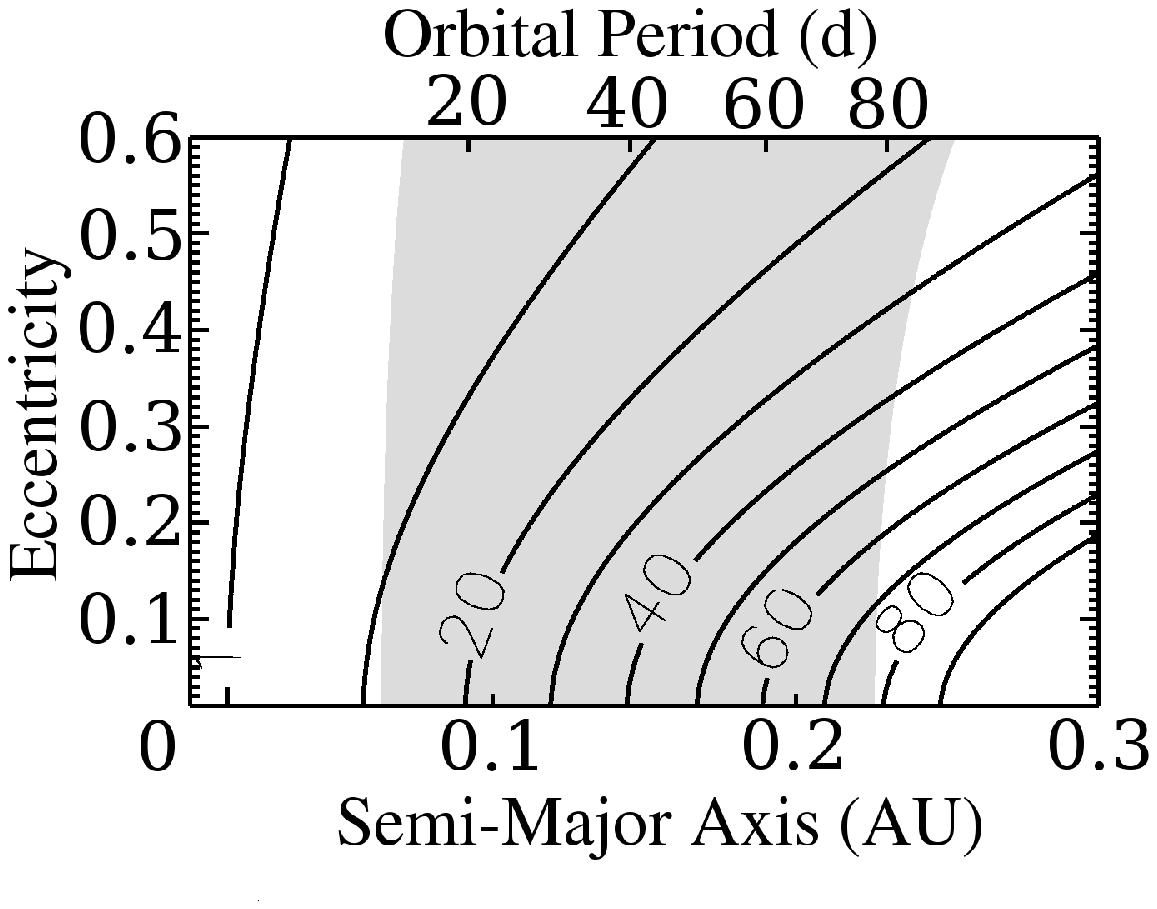}
\caption{\label{fig:rot}Contours of equilibrium rotation period in days for a 10
  $\mearth$ planet orbiting a 0.25 $\msun$ star. The gray region is the HZ
  from Barnes \etal (2008).}
\end{figure}
 
\section{Rotation Rates}
Planetary rotation rates may be modified by tidal
interactions. Although planets may form with a wide range of rotation
rates $\Omega$, tidal forces may fix $\Omega$ such that no net
exchange of rotational and orbital angular momenta occurs during one 
orbital
period. The planet is then said to be ``tidally locked,'' and the
rotation rate is ``pseudo-synchronous'' or in
equilibrium. The equilibrium rotation rate in the constant-phase-lag
model is
\begin{equation}
\label{eq:cplspin}
\Omega_{eq} = n(1 + \frac{19}{2}e^2),
\end{equation}
where $n$ is the mean motion (Goldreich 1966). Note that planets only rotate
synchronously (one side always facing the star) if $e = 0$ (the
constant-time-lag model makes the same prediction). Therefore,
the threat to habitability may have been overstated in the past, as
independently pointed out by several recent investigations (Barnes
\etal 2008; Ferraz-Mello \etal 2008; Correia \etal 2008). Figure 
\ref{fig:rot} shows the values of the equilibrium rotation period for a 
10 $\mearth$
planet orbiting a 0.25 $\msun$ star as a function of $a$ and
$e$.

\begin{figure}[t]
\plotone{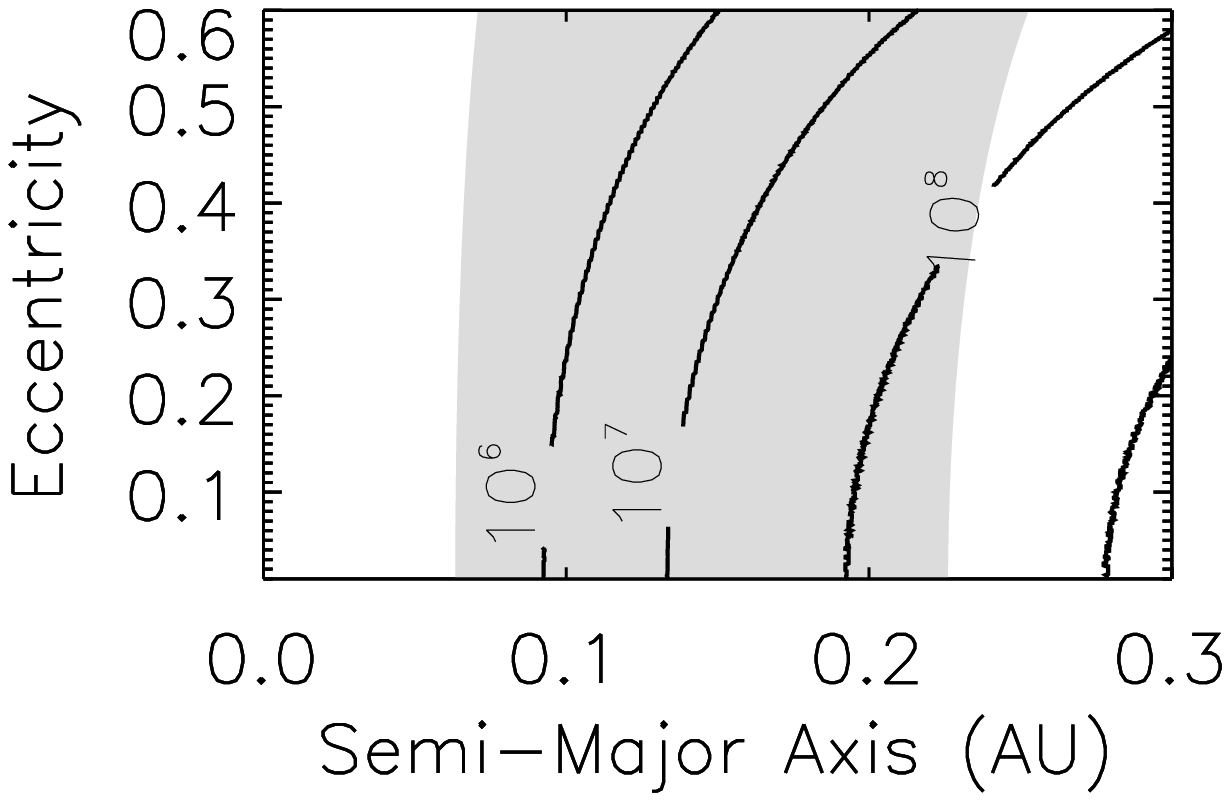}
\caption{\label{fig:obl}Time in years for a 10 $\mearth$ planet orbiting 
a 0.25 $\msun$
  star to evolve from an obliquity $\psi = 23\circdot$5 to
  $5\circdot$ The gray region is the HZ
  from Barnes \etal (2008).}
\end{figure}

\section{Obliquity}
Tidal effects tend to drive obliquities to zero or $\pi$. The constant-time-lag
model of Levrard \etal (2007) found a planet's obliquity $\psi$
changes as

\begin{equation}
  \frac{\mathrm{d}\psi}{\mathrm{d}t} = \frac{\sin{(\psi)} K_\mathrm{p}}{C_\mathrm{p} \Omega_0 n} \left( \frac{\cos{(\psi)} \epsilon_1 \Omega_0}{n} - 2 \epsilon_2 \right)
\end{equation}
where
\begin{equation}
  \epsilon_1  = \frac{1+3e^2+\frac{3}{8}e^4}{(1-e^2)^{9/2}},
\end{equation}
\begin{equation}\label{equ:K}
  K_\mathrm{p} = \frac{3}{2} k_{2,\mathrm{p}} \frac{G M_\mathrm{p}^2}{R_\mathrm{p}} \tau_\mathrm{p} n^2 \left( \frac{M_\mathrm{s}}{M_\mathrm{p}} \right)^2 \left( \frac{R_\mathrm{p}}{a} \right)^6,
\end{equation}
\begin{equation}
  C_\mathrm{p} = r_\mathrm{g,p}^2 M_\mathrm{p} R_\mathrm{p}^2,
\end{equation}
and
\begin{equation}
  \epsilon_2 = \frac{1+\frac{15}{2}e^2+\frac{45}{8}e^4+\frac{5}{16}e^6}{(1-e^2)^6}  \hspace{0.2cm}.
\end{equation}
\noindent
In the preceding equations $r_\mathrm{g,p}$ (= 0.5) is the planet's
radius of gyration (a measure of the distribution of matter inside a
body), $\Omega_0$ is the initial rotation frequency, and
$\tau_\mathrm{p}$ is the ``tidal time lag'', which in this
constant-time-lag model replaces $Q'_p$. We assumed $Q'_p = 500$ for
the planet at its initial orbital configuration and set
$\tau_\mathrm{p} = 1/(nQ'_\mathrm{p})$, \ie initially the
planet responds in the same way as in a constant-phase-lag model. In
the course of the orbital evolution, $\tau_\mathrm{p}$ was then fixed
while $n$ and $Q_\mathrm{p}$ evolved in a self-consistent system of
coupled differential equations. In Fig.~\ref{fig:obl} we show the time
for a planet with an initial obliquity of $23\circdot5$ to reach
$5^\circ$, a value which may preclude habitability (F. Selsis,
personal communication). However, obliquities may easily be modified
by other planets in the system (Atobe \etal 2004; Atobe and Ida 2007)
or a satellite (Laskar \etal 1993).

\begin{figure}[t]
\plotone{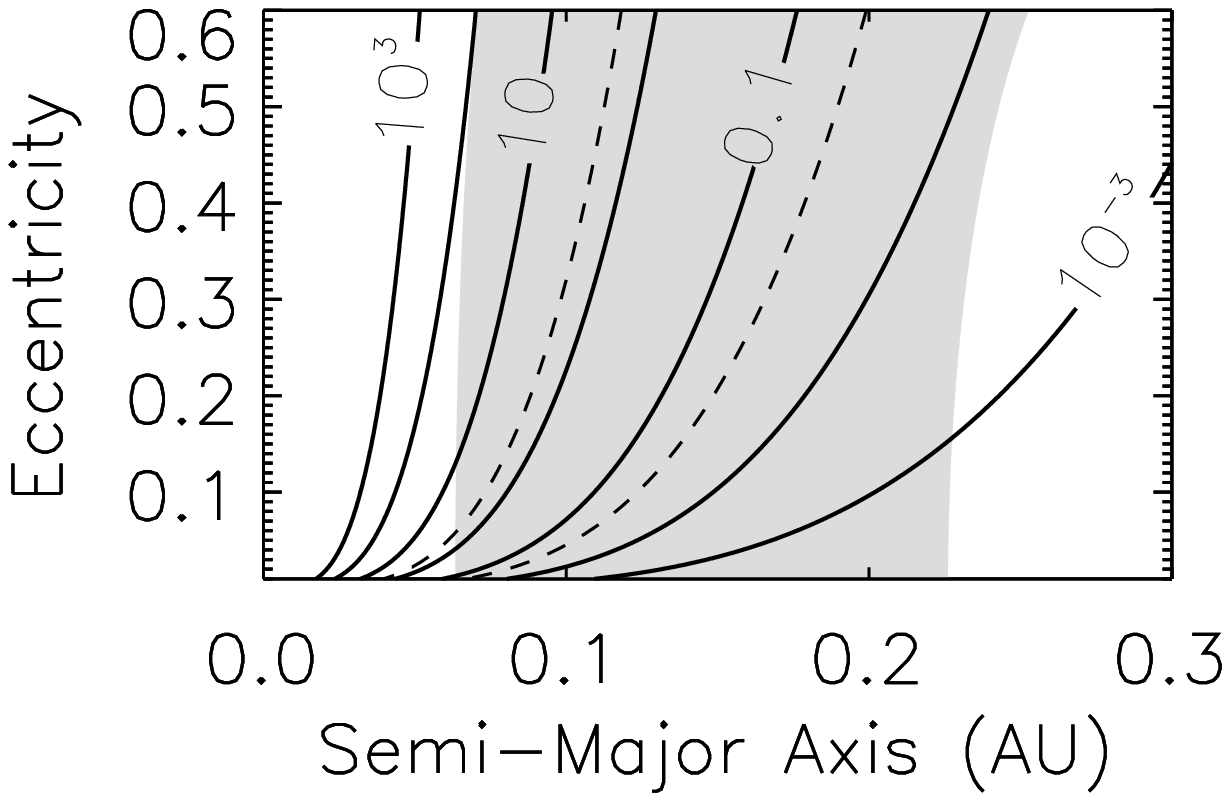}
\caption{Tidal heating fluxes for a 10 $\mearth$ planet orbiting a 0.25
  $\msun$ star. Contour labels are in W m$^{-2}$. The dashed contours
  represent the boundaries of the tidal habitable zone (Jackson \etal
  2008c; Barnes \etal 2009b). The gray region is the HZ
  from Barnes \etal (2008).\label{fig:heat}}
\end{figure}

\section{Tidal Heating}
As a body on an eccentric orbit is continually reshaped due to the varying
gravitational field, friction heats the interior. This ``tidal
heating'' is quantified in the constant-phase-lag model as
\begin{equation}
\label{eq:heat}
H = \frac{63}{4}\frac{(GM_*)^{3/2}M_*R_p^5}{Q'_p}a^{-15/2}e^2
\end{equation}
(Peale \etal 1979; Jackson 2008b). However, in order to assess the
surface effects of tidal heating on a potential biosphere, it is
customary to consider the heating flux, $h = H/4\pi R_p^2$, through the
planetary surface. Jackson \etal (2008c; see also Barnes \etal 2009b)
argued that when $h \ge 2$ W m$^{-2}$ (the value for Io [McEwen \etal 2004])
or $h \le 0.04$ W m$^{-2}$ (the limit for plate tectonics [Williams
  \etal 1997]), habitability is less likely. Barnes \etal (2009b) 
suggested
that these limits represent a ``tidal habitable zone''. In Fig.~\ref{fig:heat}
contours of tidal heating are shown for a 10 $\mearth$ planet orbiting
a 0.25 $\msun$ star. The tidal habitable zone is the region between
the dashed curves. Note that $a$ and $e$ evolve as prescribed by
Eqs.~(\ref{eq:adot} -- \ref{eq:edot}), and hence the heating fluxes
evolve with time as well.

\acknowledgements %%% Text of acknowledgements runs on after this 
RB and SNR acknowledge funding from NASA Astrobiology Institute's
Virtual Planetary Laboratory lead team, supported by NASA under
Cooperative Agreement No. NNH05ZDA001C. RG acknowledges support from
NASA's Planetary Geology and Geophysics program, grant
No. NNG05GH65G. BJ is funded by an NPP administered by ORAU. RH
is supported by a Ph.D. scholarship of the DFG Graduiertenkolleg 1351
``Extrasolar Planets and their Host Stars''.

\end{document}